\documentclass[10pt,conference]{IEEEtran}
\usepackage[utf8]{inputenc}
\usepackage{listings}
\usepackage{xcolor}
\usepackage{multirow}
\usepackage{ifthen}
\usepackage{paralist}
\usepackage{numprint}
\usepackage{graphicx}
\usepackage{xspace}

\usepackage{dblfloatfix}
\usepackage{hyperref}

\definecolor{codegreen}{rgb}{0,0.6,0}
\definecolor{codegray}{rgb}{0.5,0.5,0.5}
\definecolor{codepurple}{rgb}{0.58,0,0.82}
\definecolor{backcolour}{rgb}{0.98,0.98,0.95}

\lstdefinestyle{mystyle}{
    backgroundcolor=\color{backcolour},   
    commentstyle=\color{codegreen},
    keywordstyle=\color{codepurple},
    numberstyle=\tiny\color{codegray},
    stringstyle=\color{codepurple},
    basicstyle=\ttfamily\footnotesize,
    breakatwhitespace=false,         
    breaklines=true,                 
    captionpos=b,                    
    keepspaces=true,                 
    numbersep=5pt,                  
    showspaces=false,                
    showstringspaces=false,
    showtabs=false,                  
    tabsize=2,
    escapeinside=\%\%,
}

\title{Mining Software Repositories with a Collaborative Heuristic Repository}

\author{

\IEEEauthorblockN{Hlib Babii}
\IEEEauthorblockA{\textit{Free University of Bozen-Bolzano, Italy}} 
\and

\IEEEauthorblockN{Julian Aron Prenner}
\IEEEauthorblockA{\textit{Free University of Bozen-Bolzano, Italy}} 
\and
\IEEEauthorblockN{Laurin Stricker}
\IEEEauthorblockA{\textit{Free University of Bozen-Bolzano, Italy}} 
\and
\IEEEauthorblockN{Anjan Karmakar}
\IEEEauthorblockA{\textit{Free University of Bozen-Bolzano, Italy}} 
\and

\IEEEauthorblockN{Andrea Janes}
\IEEEauthorblockA{\textit{Free University of Bozen-Bolzano, Italy}} 
\and

\IEEEauthorblockN{Romain Robbes}
\IEEEauthorblockA{\textit{Free University of Bozen-Bolzano, Italy}} 
}

\date{October 2020}

\newboolean{showcomments}
\setboolean{showcomments}{true}
\ifthenelse{\boolean{showcomments}}{%
   \newcommand{\maybecomment}[1]{ {\lbrack #1 \rbrack} }
   }{
   \newcommand{\maybecomment}[1]{ }
}

\newcommand{\buggy}[0]{\texttt{BUG-FIXING}}
\newcommand{\bugless}[0]{\texttt{NOT-BUG-FIXING}}

\newcommand{\etal}[0]{\textit{et al.}\xspace}

\newcommand{\kword}[1]{\texttt{#1}}

\begin{document}

\maketitle

\begin{abstract}
    Many software engineering studies or tasks rely on categorizing software engineering artifacts. In practice, this is done either by defining simple but often imprecise heuristics, or by manual labelling of the artifacts. Unfortunately, errors in these categorizations impact the tasks that rely on them. To improve the precision of these categorizations, we propose to gather heuristics in a collaborative \emph{heuristic repository}, to which researchers can contribute a large amount of diverse heuristics for a variety of tasks on a variety of SE artifacts. These heuristics are then leveraged by state-of-the-art weak supervision techniques to train high-quality classifiers, thus improving the categorizations. We present an initial version of the heuristic repository, which we applied to the concrete task of commit classification. 
\end{abstract}

\section{Introduction}

Mining Software Repositories (MSR) uses data from software repositories (e.g., GitHub, Stack Overflow) to perform empirical studies and provide actionable advice to developers and managers, or to train machine learning models to provide recommendations. MSR has seen very high interest from the research community due to the availability of open-source repositories and applications in the industry. Use cases include change recommendation, bug triage, bug localisation, defect prediction, or program repair, among many others. 

For many of these tasks, an important preprocessing step is to select the subset of artifacts of interests. To take a single example, many studies are interested in changes to software systems that are correcting defects. These bug fixes are useful for tasks such as empirical studies of their characteristics, defect prediction, or program repair, to name a few. Unfortunately, selecting artifacts is not a trivial task. The data used in MSR studies is often very noisy. The primary reason for this is that developers use tools that happen to generate useful data for software researchers as a side effect; developers care about getting the job done, not easing the life of researchers. 

For source code changes for instance, they are not, for the most part, explicitly classified as bug fixes or not. In practice, researchers have to develop \emph{heuristics} to identify bug fixes, which often consist in searching for a set of keywords in commit messages. These heuristics are necessarily limited. A manual validation of such a heuristic found that it had a 36\% false positive rate, and an 11\% false negative rate \cite{berger2019impact}. This is without considering additional data quality issues in commit messages, such as the fact that up to 14\% percent of commit messages can be empty \cite{dyer2013boa}, or that commit messages may be in other languages than English, cases in which a keyword-based heuristic would fail to produce results. Moreover, up to 15\% of bug fixes can be tangled \cite{herzig2013impact}, or contain other unrelated changes such as non-essential changes \cite{kawrykow2011non}, or refactorings \cite{hora2018assessing}. The alternative to using heuristics is manual labelling of SE artifacts, which does not scale beyond a few hundreds or thousands of artifacts without significant effort \cite{robbes2019leveraging}.

Such imprecisions in preprocessing steps can compound each other. The SZZ algorithm is a heuristic that takes as input a bug fixing commit (BFC) and uses the change history to find the commit that introduced the bug, or bug-introducing commit (BIC) \cite{sliwerski2005changes}. A precise algorithm to identify BICs from a data corpus would be very valuable to train machine learning models to precisely classify changes as clean or defective \cite{kim2008classifying}, which would be very useful during code review. Unfortunately, SZZ is affected by many of the issues described above as it builds on previous heuristics; a manual investigation reported the F-score of SZZ at  42 to 44\% on two datasets \cite{rodriguez2018towards}; this impacts change-based defect prediction models \cite{rodriguez2020bugs}.

While we focus on bug fix identification as a single, running example, this problem is more general, spanning any classification of interest (e.g., tangled commits \cite{herzig2013impact}), on any artifact of interest (e.g., bug reports \cite{herzig2013s}, Stack Overflow \cite{rahman2019cleaning}), for which simplistic heuristics are used to categorize the artifacts. We argue that \emph{using heuristics is not the problem, rather it is using not enough of them, for lack of a way to effectively gather, integrate, and combine them}.

Instead, we propose a two-pronged approach:
\begin{inparaenum}[1)]
\item build a collaborative, open repository of heuristics, in order to gather an exhaustive registry of SE artifact types, their categorization tasks, and associated heuristics and
\item use state of the art techniques to combine these heuristics, in order to increase their reliability when categorizing SE artifacts in a categorization task. We instantiated this approach for the task of classifying commits as bug fixing or not, using a variety of heuristics to do so.
\end{inparaenum}
A preliminary evaluation shows that the approach already exceeds the accuracy of the baseline heuristic; we expect the accuracy to improve further by adding additional heuristics, and have plans to apply the approach to additional tasks. Thus, we invite researchers to define new tasks and contribute heuristics to our heuristic repository.

\section{Combining Heuristics with Snorkel}

\textbf{Snorkel in a nutshell.} We use the Snorkel framework \cite{ratner2017snorkel} to define and combine heuristics, as a form of \emph{weak supervision}. Weak supervision replaces small, hand-annotated datasets with larger but noisier datasets annotated by heuristics. 

This is intuitively what MSR researchers do now, except that Snorkel supports combining a \emph{large number of varied heuristics}, instead of being limited to a small number of keywords. 
Snorkel represents heuristics as labelling functions \cite{ratner2017snorkel}, which are short programs that can annotate data with noisy labels. Apart from simple pattern matching, e.g. looking for keywords in a commit message \cite{casalnuovo2017gitcproc}, labelling functions can apply more complex logic to single data points (e.g. known bug patterns \cite{karampatsis2020often}). Furthermore, labelling functions can use information sources other than the artifact itself:
\begin{itemize}
    \item external data (e.g., a bug report linked to a commit);
    \item an existing hand-annotated dataset;
    \item an existing low-quality classifier for the task. 
\end{itemize}

These labelling functions have several important characteristics: 1) they can have limited coverage on the data, and can \emph{abstain} to annotate a given piece of data (e.g., label a commit message as \buggy{} if it contains a keyword, but do not necessarily annotate it as \bugless{} if it does not), 2) since they are heuristics, they are assumed to be imperfect and noisy, 3) they can be correlated (e.g., keywords \kword{bug} and \kword{fix} in ``fix bug \#1234'')  or contradict each other (e.g., keywords \kword{fix} and \kword{typo} in ``fix a small typo'').

Snorkel then leverages the labelling functions to learn a generative model of the labels, which it can use to label previously unlabelled data with probabilistic labels. The generative model learns which labelling functions are correlated, or conflicting, and learns to estimate their accuracy; in short, it learns to optimally combine the heuristics to label data. This is an unsupervised process that does not require previously labelled data as ground truth. This \emph{label model} can then annotate datasets with probabilistic labels (e.g., the commit ``fix bug \#1234'' is 90\% likely to be \buggy{}, while  ``fix a small typo'' is 70\% likely to be \bugless{}). The label model is more accurate than simply modelling each labelling function as a ``vote`` and picking the majority class. Snorkel gracefully handles noise: while less noisy heuristics are obviously better, the only requirement for a heuristic to improve performance is to perform better than chance \cite{ratner2017snorkel}.

\textbf{Training classifiers.} In essence, the label model acts as a kind of classifier. However, it can also be used to label data for a more powerful classifier such as a deep neural network. These classifiers significantly outperform classifiers trained on a smaller amount of hand labelled data; they usually outperform the label model itself, as they can learn from additional correlations in the data, and generalize to situations not covered by the heuristics. Note that the heuristics can leverage external knowledge to label the data, even if this data will be inaccessible to the end classifier. The higher quality of the resulting labelled data can still end up in improved performance. A case study at Google describes such as scenario \cite{bach2019snorkel} involving data sources that are not usable in production (e.g. data that is too slow or too sensitive to access).

\textbf{Task redefinition.} Finally, a key advantage of Snorkel's approach, with respect to hand-labelled data, is that it makes it much easier to redefine a task if needed. For instance, if a binary classification task changes to 3-way classification, it is much easier to add or redefine labelling functions, rather than relabelling the entire dataset from scratch. 



\section{BOHR: The Big Old Heuristic Repository}

While Snorkel on its own is ideal for a small team working on a single task, we argue that the data quality issues facing the MSR community call for something more. The MSR community is a large group of researchers, working on a very wide variety of tasks spanning from a set of possible SE artifacts. While each of these tasks can be addressed in isolation, this is a missed opportunity. For instance, several tasks rely on distinguishing bug fixes, refactorings, or tangled changes from other commits; all would benefit from a consolidated effort at improving this preprocessing step. 

\textbf{The repository}. To facilitate the sharing and reuse of knowledge necessary for the community to progress on these tasks, we are building BOHR (the ``Big Old Heuristic Repository''), a repository of SE artifact labelling heuristics and associated classification tasks. The goals of BOHR are to:
\begin{inparaenum}[1)]
\item increase the visibility of heuristics by grouping them in a single location;
\item ease their definition and use by abstracting away their application to SE artifacts;
\item ease the reuse of heuristics for new tasks and artifacts,
\item ease contributions by lowering barriers to contribute to the repository; and
\item ease the generation of label models and datasets to train an end classifier, and ensure their reproducibility.
\end{inparaenum}
We are in the process of instantiating this vision: for now, BOHR supports two types of software engineering artifacts, commits and issue reports, and we have applied it to commit classification tasks.


\textbf{Defining heuristics.} 
BOHR facilitates the definition and application of heuristics to SE artifacts, by converting them to homogeneous objects with immediately accessible structure. Heuristics take these objects as input, and thus can easily hone in on properties of interest (Listing \ref{lst:heuristic-example}). Moreover, BOHR provides links between different types of artifacts, e.g. a heuristic applied to a commit can access the bug report linked to it. Common types of heuristics are abstracted away: e.g., adding a keyword only requires modifying a keyword list in a text file; a heuristic that relies on a common external tool can use its output without knowing how to run the tool. 

\begin{lstlisting}[language=Python, caption={Example of heuristic applied to a commit artifact}, label={lst:heuristic-example}]
@Heuristic(Commit)
def bugless_if_many_files_changed(commit: Commit):
    return !BugFix if len(commit.files)>6 else None
\end{lstlisting}

\begin{figure}
    \centering
    \includegraphics[width=0.41\textwidth]{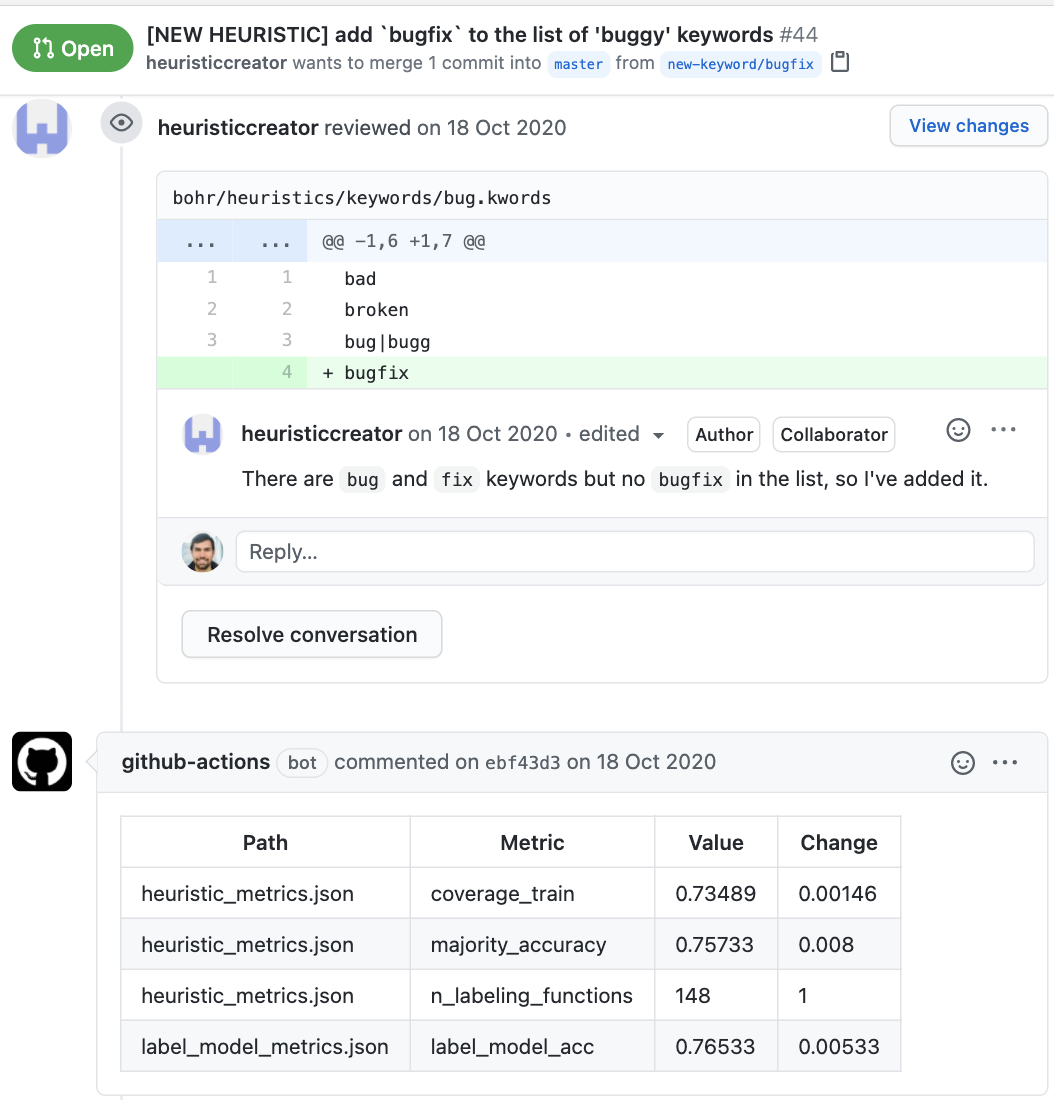}
    \caption{Pull request example}
    \label{fig:pr}
\end{figure}

\textbf{Contributing heuristics.} To lower barriers to entry, BOHR is hosted on github (\texttt{\url{https://github.com/giganticode/bohr}}), and is thus accessible to the research community at large. Further, BOHR's repository also makes it easy for researchers to contribute new heuristics. A researcher that wants to contribute a new heuristic can simply fork the repository, implement the heuristic, and submit a pull request (PR). The PR will trigger a bot that re-runs the label model and posts informative metrics on the label model's performance as comments to the PR, as shown in Figure \ref{fig:pr}. For the simplest cases (e.g., translating keywords in languages other than English to increase the coverage to commit messages in other languages), it might not even be necessary to install anything before contributing. To make sure that heuristics developed for a specific dataset can be reused and generalize well for other datasets, metrics are calculated on a stand-alone test-set(s). The PR bot provides insights into the newly developed heuristic, such as its coverage of the dataset, how much it overlaps and conflicts with other heuristics, and what is its individual contribution to the overall increase of accuracy.

Once a label model is created for a given task, BOHR can facilitate the generation of the final dataset (e.g., selecting the attributes of interest to include in the dataset), which can then be used as-is, or passed to an end classifier for training.

\textbf{Defining tasks and task variants.} While Snorkel labelling functions directly return a label for the task at hand, we are currently adding an additional layer to separate heuristics from concrete tasks. This will allow to ease the definition of new tasks, by reusing existing heuristics. Another use case for this is the definition of task variants: for some tasks, the inclusion criteria may vary (e.g., some works do not consider bug fixes in test code as bugs \cite{zafar2019towards}). We consider this a significant advantage of the approach: instead of implicitly encoding these decisions in the dataset, they will be explicitly encoded as heuristics. Researchers will be able to decide whether or not to include a specific decision in the set of heuristics they use for a task (e.g., they can choose to consider fixes to test code as bug fixes). This benefit is in addition to the effort saved by avoiding to relabel a dataset when the task definition changes. 

\section{Example: Commit Classification}

As a first task, we defined a commit classification task with BOHR. We classify whether commits are bug fixes (\buggy{} class), or other types of commits (\bugless{} class). 
This classification is very often done using ad-hoc, simple keyword-based heuristics, as is the case in several studies \cite{kim2008classifying, ray2014large, karampatsis2020often, tufano2019empirical}. These heuristics look for keywords indicating bugs, and assume that commit messages not matching any keyword are not bug fixes.

\textbf{Heuristics used.} In this first task, we principally employ keyword-based heuristics as well, reserving an investigation of other heuristic sources for future work. In contrast to previous work, however, our heuristics cover a much broader set of keywords (several dozens), which cover both classes. Our heuristics also leverage Snorkel's ability to use external resources for labelling, that an end classifier may not have access to. In particular, we keep track of links between commits and issue reports, and we also define heuristics that look for keywords in issue labels and issue contents. Since less than 10\% of commits have such links in practice, relying on issues in an end classifier would be problematic.

\textbf{Datasets.} To build a label model and a end classifier, we build a dataset of \numprint{200000} commits gathered from Github via the Github API. Our dataset has \numprint{68000} commits for which we have a link to an issue; the remaining commits do not have such links, so that both type of commits are represented in the dataset. To evaluate the accuracy of our label model and end classifier for commit classification, we use three existing, hand-labelled datasets as test sets, shared by three different studies \cite{herzig2013s, levin2017boosting, berger2019impact}. Of note, Levin et al. also train a classifier \cite{levin2017boosting},  however, we do not compare to it directly since they use cross-validation on their hand-labelled dataset for training and validation, while we use it as an unseen test set. The dataset of Herzig and Zeller is a dataset of issue reports that was manually classified, rather than commits; to use it as a test set, we searched for all the commits that linked to these issue reports. We use these commits, not the original issue reports, as our test set. 



\begin{table}
  \footnotesize
  \setlength\extrarowheight{1pt}
  \setlength{\tabcolsep}{2pt}

  \label{tab:performance}
  \caption{Accuracy, macro F1 and per-class precision and recall for different models and datasets.} 
  
  \begin{center}
    \begin{tabular}{@{\extracolsep{3pt}}llllllll@{}}
       \multirow{2}{*}{\textbf{Dataset}} & 
       \multirow{2}{*}{\textbf{Model}} &  \multirow{2}{*}{\textbf{Acc.}} & \multirow{2}{*}{\textbf{F1}} & \multicolumn{2}{c}{\textbf{Precision}} & \multicolumn{2}{c}{\textbf{Recall}} \\
       \cline{5-6}
       \cline{7-8}
       &   &   &   &  \textbf{bugless} & \textbf{bug} & \textbf{bugless} & \textbf{bug} \\
       \parbox[t]{2mm}{\multirow{4}{*}{\rotatebox[origin=c]{90}{Levin~\cite{levin2017boosting}}}} 
       & Tufano et al. & 0.651 & 0.557 & 0.621 & \textbf{0.902} & \textbf{0.982} & 0.220 \\    
       & GitCProc & 0.764 & 0.748 & 0.741 & 0.813 & 0.896 & 0.592\\
       & Label Model & \underline{0.803} & \underline{0.798} & \textbf{0.815} & 0.786 & 0.843 & \textbf{0.750} \\
       & Transformer & \textbf{0.825} & \textbf{0.816} & \underline{0.796} & \underline{0.882} & \underline{0.929} & \underline{0.690} \\
       \cline{1-8}

       \parbox[t]{2mm}{\multirow{4}{*}{\rotatebox[origin=c]{90}{Berger~\cite{berger2019impact}}}} 
       & Tufano et al. & 0.720 & 0.629 & 0.699 & \textbf{0.857} & \textbf{0.970} & 0.300 \\       
       & GitCProc & \underline{0.787} & \underline{0.773} & 0.835 & 0.708 & \underline{0.821} & 0.729 \\
       & Label Model & 0.773 & 0.769 & \textbf{0.891} & 0.650 & 0.728 & \textbf{0.850} \\
       & Transformer & \textbf{0.800} & \textbf{0.790} & \underline{0.860} & \underline{0.712} & 0.813 & \underline{0.779} \\
       \cline{1-8}

       \parbox[t]{2mm}{\multirow{4}{*}{\rotatebox[origin=c]{90}{Herzig~\cite{herzig2013s}}}} 
       & Tufano et al. & 0.589 & 0.531 & 0.543 & \textbf{0.877} & \textbf{0.966} & 0.232 \\
       & GitCProc & 0.654 & 0.634 & 0.594 & 0.835 & \underline{0.915} & 0.407 \\
       & Label Model & \textbf{0.720} & \textbf{0.719} & \textbf{0.677} & 0.779 & 0.810 & \textbf{0.635} \\
       & Transformer & \underline{0.708} & \underline{0.700} & \underline{0.645} & \underline{0.838} & 0.890 & \underline{0.535} \\

    \end{tabular}  
  \end{center}
\label{tab:senti-results}
\end{table}
\setlength{\tabcolsep}{6pt}


\textbf{Training a label model.} We trained our label model on \numprint{200000} commits, which took 30 minutes on a dedicated machine. The label model is able to output labels to roughly 80\% of the commits. This is because this label model is still limited: it only looks at keywords in the commit message, issue labels, and issue contents. Up to \numprint{40000}
commit messages do not match any keywords, showing that there is a margin for improvement by adding additional sources of heuristics, such as heuristics that look at source code changes in addition to the commit message, or even additional keywords: anecdotally, we also see commits that are in languages other than English in the training set. On the test datasets, the label model also \emph{abstains} to annotate between 10 and 20\% of the time; and when it does, we assign the \bugless{} label.

\textbf{Training an end classifier.} We also train an end classifier by fine-tuning a 6-layer Transformer \cite{vaswani2017attention} deep neural network, with an open vocabulary \cite{karampatsis2020big}. The model was pre-trained using the \textit{RoBERTa} training regime \cite{liu2019roberta} on Stack Overflow comments, so that its initial model weights are adapted to the Software Engineering domain \cite{robbes2019leveraging}. We then fine-tuned as a classifier on the \numprint{160000} commit messages that the Label Model could label. Unlike the label model, the trained classifier never abstains.

\textbf{Baseline.} As baseline for comparison, we use heuristics that were used in the literature. GitCProc \cite{casalnuovo2017gitcproc} is a tool that was used in several studies, such as the programming language study of Ray \etal; it matches 17 keywords to the \buggy{} class. Tufano \etal \cite{tufano2019empirical} used another heuristic, using only 6 keywords to precisely select bug fixes only.

\textbf{Results.} Table~\ref{tab:performance} shows the performance of the two baselines, as well as the label model and the Transformer end classifier. We report accuracy and F-measure, as well as precision and recall for each tasks. We can see that the heuristic used in Tufano \etal has consistently the highest precision in identifying bug fixes---as intended, but it also has a very low recall; its uneven performance cause it to consistently score low in terms of F1. On the Levin and Herzig datasets, there is a gap between the second heuristic, GitCProc, and our two approaches (the Label Model and the Transformer). On the Berger dataset, the Transformer is performing best, with GitCProc following, and the Label Model a very close third; note that this dataset is the smallest of the three. On two out of three datasets, the Transformer outperforms the Label Model, which is promising for the future; we expect that a Transformer trained on change data in addition to commit messages will perform better still, particularly since the LabelModel trained on commit messages abstains 10 to 20\% of the time.


\textbf{Potential for task variants.} We also examined the performance of individual labelling functions on the three datasets. We saw that several labelling functions had ample variations in performance on the datasets. Relatively common terms such as \texttt{test}, \texttt{patch}, \texttt{version}, or \texttt{close} were either positively or negatively associated with a commit being a bug fix, depending on the dataset. This may indicate that, when labelling each dataset, different researchers had a somewhat different definition of what a bug is. While this complicates the definition of a general classifier, this also shows the potential of an approach making heuristics explicit: if the definition of the task changes, it is much easier to change heuristics, than to relabel an entire dataset.


\section{Conclusions and Future Work}
We presented BOHR, a repository in which researchers can contribute heuristics to label SE data. BOHR uses state of the art weak supervision techniques to combine these heuristics to train classifiers operating on SE data. We have shown the potential of the approach, by applying initial heuristics to the task of commit classification. We presented very early results, and expect significant work ahead, along a few lines of work.

\textbf{Improving binary commit classification.} There are several ways to further improve the binary commit classification task. We plan to add additional heuristics that look at other data sources, such as actual change patterns, to create an improved labelling model. We also plan to define models that look at data beyond the commit message, such as change metrics or actual source code changes \cite{hoang2019patchnet}. A further way to improve performance will also be to scale up the end model, by training a larger capacity model on a larger dataset. Another possible improvement is to develop classifiers that rely less on the commit message and principally---or exclusively---consider the changes instead: many commits have messages that either are empty, uninformative, or in other languages than English.

\textbf{Extending commit classification.} We will continue our initial experiments with variants of the commit classification task. In particular, there are several valuable types of commits that are not specifically classified. These include several types of large commits \cite{hindle2008large}, commits that contain tangled changes \cite{herzig2013impact}, or refactorings \cite{hora2018assessing}, in addition to commits that are simply adding new features. 
We will develop new heuristics to detect these kinds of commits, based on commit messages, issue contents, and source code changes, thus developing several variants of the commit classification task.

\textbf{Extending BOHR and involving the community.} Developing new classification tasks and variants will be crucial to improve the useability and extensibility of BOHR. In the process, we will also extend BOHR to support more types of SE artifacts.
This will allow to use BOHR to categorize issues, discussions, documentation, app reviews, or entire projects, to name a few.
Given the amount of possible tasks and variants, we will not be able to do this alone. We would thus like to involve the community in this effort. We plan to make it possible for other researchers to add new tasks easily, and call on the community to do so.

\section{Data Availability and Aknowledgements}
Information on how to access the code, heuristics, and datasets used in this study is available at: \texttt{\url{https://github.com/giganticode/bohr/wiki/NIER-2021}}. This work was partially funded by the IDEALS and ADVERB projects, funded by the Free University of Bozen-Bolzano. Parts of the results of this work were computed on the Vienna Scientific Cluster (VSC). 








\bibliographystyle{IEEEtran}
\bibliography{bibliography}

\begin{thebibliography}{10}
\providecommand{\url}[1]{#1}
\csname url@samestyle\endcsname
\providecommand{\newblock}{\relax}
\providecommand{\bibinfo}[2]{#2}
\providecommand{\BIBentrySTDinterwordspacing}{\spaceskip=0pt\relax}
\providecommand{\BIBentryALTinterwordstretchfactor}{4}
\providecommand{\BIBentryALTinterwordspacing}{\spaceskip=\fontdimen2\font plus
\BIBentryALTinterwordstretchfactor\fontdimen3\font minus
  \fontdimen4\font\relax}
\providecommand{\BIBforeignlanguage}[2]{{%
\expandafter\ifx\csname l@#1\endcsname\relax
\typeout{** WARNING: IEEEtran.bst: No hyphenation pattern has been}%
\typeout{** loaded for the language `#1'. Using the pattern for}%
\typeout{** the default language instead.}%
\else
\language=\csname l@#1\endcsname
\fi
#2}}
\providecommand{\BIBdecl}{\relax}
\BIBdecl

\bibitem{berger2019impact}
E.~D. Berger, C.~Hollenbeck, P.~Maj, O.~Vitek, and J.~Vitek, ``On the impact of
  programming languages on code quality: a reproduction study,'' \emph{ACM
  Transactions on Programming Languages and Systems (TOPLAS)}, vol.~41, no.~4,
  pp. 1--24, 2019.

\bibitem{dyer2013boa}
R.~Dyer, H.~A. Nguyen, H.~Rajan, and T.~N. Nguyen, ``Boa: A language and
  infrastructure for analyzing ultra-large-scale software repositories,'' in
  \emph{2013 35th International Conference on Software Engineering
  (ICSE)}.\hskip 1em plus 0.5em minus 0.4em\relax IEEE, 2013, pp. 422--431.

\bibitem{herzig2013impact}
K.~Herzig and A.~Zeller, ``The impact of tangled code changes,'' in \emph{2013
  10th Working Conference on Mining Software Repositories (MSR)}.\hskip 1em
  plus 0.5em minus 0.4em\relax IEEE, 2013, pp. 121--130.

\bibitem{kawrykow2011non}
D.~Kawrykow and M.~P. Robillard, ``Non-essential changes in version
  histories,'' in \emph{2011 33rd International Conference on Software
  Engineering (ICSE)}.\hskip 1em plus 0.5em minus 0.4em\relax IEEE, 2011, pp.
  351--360.

\bibitem{hora2018assessing}
A.~Hora, D.~Silva, M.~T. Valente, and R.~Robbes, ``Assessing the threat of
  untracked changes in software evolution,'' in \emph{Proceedings of the 40th
  International Conference on Software Engineering}, 2018, pp. 1102--1113.

\bibitem{robbes2019leveraging}
R.~Robbes and A.~Janes, ``Leveraging small software engineering data sets with
  pre-trained neural networks,'' in \emph{2019 IEEE/ACM 41st International
  Conference on Software Engineering: New Ideas and Emerging Results
  (ICSE-NIER)}.\hskip 1em plus 0.5em minus 0.4em\relax IEEE, 2019, pp. 29--32.

\bibitem{sliwerski2005changes}
J.~{\'S}liwerski, T.~Zimmermann, and A.~Zeller, ``When do changes induce
  fixes?'' \emph{ACM sigsoft software engineering notes}, vol.~30, no.~4, pp.
  1--5, 2005.

\bibitem{kim2008classifying}
S.~Kim, E.~J. Whitehead, and Y.~Zhang, ``Classifying software changes: Clean or
  buggy?'' \emph{IEEE Transactions on Software Engineering}, vol.~34, no.~2,
  pp. 181--196, 2008.

\bibitem{rodriguez2018towards}
G.~Rodr{\'\i}guez~P{\'e}rez \emph{et~al.}, ``Towards an empirical model to
  identify when bugs are introduced,'' 2018.

\bibitem{rodriguez2020bugs}
G.~Rodr{\'\i}guez-P{\'e}rez, G.~Robles, A.~Serebrenik, A.~Zaidman, D.~M.
  Germ{\'a}n, and J.~M. Gonzalez-Barahona, ``How bugs are born: a model to
  identify how bugs are introduced in software components,'' \emph{Empirical
  Software Engineering}, pp. 1--47, 2020.

\bibitem{herzig2013s}
K.~Herzig, S.~Just, and A.~Zeller, ``It's not a bug, it's a feature: how
  misclassification impacts bug prediction,'' in \emph{2013 35th International
  Conference on Software Engineering (ICSE)}.\hskip 1em plus 0.5em minus
  0.4em\relax IEEE, 2013, pp. 392--401.

\bibitem{rahman2019cleaning}
M.~Rahman, P.~Rigby, D.~Palani, and T.~Nguyen, ``Cleaning stackoverflow for
  machine translation,'' in \emph{2019 IEEE/ACM 16th International Conference
  on Mining Software Repositories (MSR)}.\hskip 1em plus 0.5em minus
  0.4em\relax IEEE, 2019, pp. 79--83.

\bibitem{ratner2017snorkel}
A.~Ratner, S.~H. Bach, H.~Ehrenberg, J.~Fries, S.~Wu, and C.~R{\'e}, ``Snorkel:
  Rapid training data creation with weak supervision,'' in \emph{Proceedings of
  the VLDB Endowment. International Conference on Very Large Data Bases},
  vol.~11, no.~3.\hskip 1em plus 0.5em minus 0.4em\relax NIH Public Access,
  2017, p. 269.

\bibitem{casalnuovo2017gitcproc}
C.~Casalnuovo, Y.~Suchak, B.~Ray, and C.~Rubio-Gonz{\'a}lez, ``Gitcproc: a tool
  for processing and classifying github commits,'' in \emph{Proceedings of the
  26th ACM SIGSOFT International Symposium on Software Testing and Analysis},
  2017, pp. 396--399.

\bibitem{karampatsis2020often}
R.-M. Karampatsis and C.~Sutton, ``How often do single-statement bugs occur?
  the manysstubs4j dataset,'' in \emph{Proceedings of the 17th International
  Conference on Mining Software Repositories}, 2020, pp. 573--577.

\bibitem{bach2019snorkel}
S.~H. Bach, D.~Rodriguez, Y.~Liu, C.~Luo, H.~Shao, C.~Xia, S.~Sen, A.~Ratner,
  B.~Hancock, H.~Alborzi \emph{et~al.}, ``Snorkel drybell: A case study in
  deploying weak supervision at industrial scale,'' in \emph{Proceedings of the
  2019 International Conference on Management of Data}, 2019, pp. 362--375.

\bibitem{zafar2019towards}
S.~Zafar, M.~Z. Malik, and G.~S. Walia, ``Towards standardizing and improving
  classification of bug-fix commits,'' in \emph{2019 ACM/IEEE International
  Symposium on Empirical Software Engineering and Measurement (ESEM)}.\hskip
  1em plus 0.5em minus 0.4em\relax IEEE, 2019, pp. 1--6.

\bibitem{ray2014large}
B.~Ray, D.~Posnett, V.~Filkov, and P.~Devanbu, ``A large scale study of
  programming languages and code quality in github,'' in \emph{Proceedings of
  the 22nd ACM SIGSOFT International Symposium on Foundations of Software
  Engineering}, 2014, pp. 155--165.

\bibitem{tufano2019empirical}
M.~Tufano, C.~Watson, G.~Bavota, M.~D. Penta, M.~White, and D.~Poshyvanyk, ``An
  empirical study on learning bug-fixing patches in the wild via neural machine
  translation,'' \emph{ACM Transactions on Software Engineering and Methodology
  (TOSEM)}, vol.~28, no.~4, pp. 1--29, 2019.

\bibitem{levin2017boosting}
S.~Levin and A.~Yehudai, ``Boosting automatic commit classification into
  maintenance activities by utilizing source code changes,'' in
  \emph{Proceedings of the 13th International Conference on Predictive Models
  and Data Analytics in Software Engineering}, 2017, pp. 97--106.

\bibitem{vaswani2017attention}
A.~Vaswani, N.~Shazeer, N.~Parmar, J.~Uszkoreit, L.~Jones, A.~N. Gomez,
  {\L}.~Kaiser, and I.~Polosukhin, ``Attention is all you need,'' in
  \emph{Advances in neural information processing systems}, 2017, pp.
  5998--6008.

\bibitem{karampatsis2020big}
R.-M. Karampatsis, H.~Babii, R.~Robbes, C.~Sutton, and A.~Janes, ``Big code!=
  big vocabulary: Open-vocabulary models for source code,'' \emph{arXiv
  preprint arXiv:2003.07914}, 2020.

\bibitem{liu2019roberta}
Y.~Liu, M.~Ott, N.~Goyal, J.~Du, M.~Joshi, D.~Chen, O.~Levy, M.~Lewis,
  L.~Zettlemoyer, and V.~Stoyanov, ``Roberta: A robustly optimized bert
  pretraining approach,'' \emph{arXiv preprint arXiv:1907.11692}, 2019.

\bibitem{hoang2019patchnet}
T.~Hoang, J.~Lawall, Y.~Tian, R.~J. Oentaryo, and D.~Lo, ``Patchnet:
  Hierarchical deep learning-based stable patch identification for the linux
  kernel,'' \emph{IEEE Transactions on Software Engineering}, 2019.

\bibitem{hindle2008large}
A.~Hindle, D.~M. German, and R.~Holt, ``What do large commits tell us? a
  taxonomical study of large commits,'' in \emph{Proceedings of the 2008
  international working conference on Mining software repositories}, 2008, pp.
  99--108.

\end{thebibliography}

\end{document}